\documentclass[journal]{IEEEtran}

\usepackage{graphicx,multicol,subfigure}
\usepackage{psfrag,amsbsy,pstricks,amsmath,amsfonts,amssymb,cite}
\usepackage{algorithm,algpseudocode}
\usepackage{epstopdf}
\usepackage{bbm}
\usepackage{color}
\usepackage{amsmath}
\usepackage{subfigure}

\usepackage{booktabs}

\newcommand{\ty}{\text{y}}
\newcommand{\tD}{\text{D}}

\newcommand{\tby}{\textbf{y}}
\newcommand{\tbD}{\textbf{D}}

\newcommand{\h}{\boldsymbol{h}}

\newcommand{\bx}{\boldsymbol{x}}
\newcommand{\by}{\boldsymbol{y}}
\newcommand{\bh}{\boldsymbol{h}}

\newcommand{\bz}{\boldsymbol{z}}
\newcommand{\bn}{\boldsymbol{n}}

\newcommand{\bdelta}{\boldsymbol{\delta}}

\newcommand{\bX}{\boldsymbol{X}}
\newcommand{\bF}{\boldsymbol{F}}

\newcommand{\mP}{\mathcal{P}}

\newcommand{\mO}{\mathcal{O}}

\newcommand{\tra}{^{\textrm{T}}}

\newcommand{\CN}{\mathcal{CN}}
\newcommand{\Ga}{\mathfrak{Ga}}
\newcommand{\Be}{\mathfrak{Be}}

\newcommand{\balpha}{\boldsymbol{\alpha}}

\newcommand{\bgamma}{\boldsymbol{\gamma}}
\newcommand{\bpi}{\boldsymbol{\pi}}

\begin{document}
%
\title{Low-Complexity Message Passing Based Massive MIMO Channel Estimation by Exploiting Unknown Sparse Common Support with Dirichlet Process}
\author{Zhengdao~Yuan, Chuanzong~Zhang, Zhongyong Wang and Qinghua Guo %
\thanks{This work is supported by the National Natural Science
Foundation of China (NSFC 61172086, NSFC U1204607, NSFC 61201251).}
\thanks{Z. Yuan is with the National Digital Switching System Engineering and Technological Research and Development Center, and the Zhengzhou Institute of Information Science and Technology, Zhengzhou 450001, China (e-mail: yuan\_zhengdao@foxmail.com).}
\thanks{C. Zhang is with the School of Information Engineering, Zhengzhou University, Zhengzhou 450001, China, and the Department of Electronic Systems, Aalborg University, Aalborg 9220, Denmark (e-mail: ieczzhang@gmail.com).}
\thanks{Z. Wang is with the School of Information Engineering, Zhengzhou University, Zhengzhou 450001, China (e-mail: iezywang@zzu.edu.cn).}
\thanks{Q. Guo is with the School of Electrical, Computer and Telecommunications Engineering, University of Wollongong, Wollongong, NSW 2522, Australia, and also with the School of Electrical, Electronic and Computer Engineering, University of Western Australia, Crawley, WA 6009, Australia (e-mail: qguo@uow.edu.au).}}
\maketitle

\begin{abstract}
This paper investigates the problem of estimating sparse channels in massive MIMO systems. Most wireless channels are sparse with large delay spread, while some channels can be observed having sparse common support (SCS) within a certain area of the antenna array, i.e., the antenna array can be grouped into several clusters according to the sparse supports of channels. The SCS property is attractive when it comes to the estimation of large number of channels in massive MIMO systems. Using the SCS of channels, one expects better performance, but the number of clusters and the elements for each cluster are always unknown in the receiver.
In this paper, {the Dirichlet process} is exploited to model such sparse channels where those in each cluster have SCS. We proposed a low complexity message passing based sparse Bayesian learning to perform channel estimation in massive MIMO systems by using combined BP with MF on a factor graph. Simulation results demonstrate that the proposed massive MIMO sparse channel estimation outperforms the state-of-the-art algorithms. Especially, it even shows better performance than the variational Bayesian method applied for massive MIMO channel estimation.\end{abstract}


\section{Introduction}\label{Sec:intro}
Deploying of multiple antennas for wireless communication systems often yield significant advantages on the performance of power gain, channel robustness, diversity and spatial multiplexing \cite{Rusek2013}, \cite{Larsson2014}. Therefor multiple-input-multiple-output (MIMO) technology has already attracted widespread attention of researchers \cite{Rusek2013,Larsson2014,Shariati2014,Wen2015,Marzetta2006,Wu2016,Masood2015}. However, accurate channel estimation is needed to realizing the full potential of MIMO systems \cite{Shariati2014}, \cite{Wen2015}. With the number of transmit antennas increasing, the receiver have to estimate proportionally more channel coefficients, which in turn increases the pilot overhead and tends to reduce the overall MIMO throughput gains \cite{Marzetta2006}. Hence, exploring efficient channel estimation technology for massive MIMO systems, required less computational complexity and number of pilots, is a challenge, which has been thoroughly addressed in \cite{Wen2015,Marzetta2006,Wu2016,Masood2015}.

To reduce the overhead, some works studied and exploited \textit{sparse common support} (SCS) approximately existing in the sparse channels of MIMO systems \cite{Barbotin2011,Masood2015,Prasad2015}. It is reasonable to assume that the antennas closely arranged will observe almost the same echoes from different reflectors or scatterers, and therefore the corresponding sparse channels will exhibit common support.
\cite{Barbotin2011} expounds that two channel taps are resolvable if the time difference of arrival is larger than $\frac{1}{10{Bw}}$, where $Bw$ is the signal bandwidth. In the other words, the channels corresponding to two antennas have SCS, when their distance is less than $\frac{C}{10Bw}$ with $C$ standing for the speed of light. The channel estimation algorithms \cite{Barbotin2011,Prasad2015} exploiting SCS property for all the channels perform well with less pilot overhead in the case of the farthest antennas of an array close enough. Masood \textit{et al.} studied the SCS property for different antenna arrays with several typical communication standards \cite{Masood2015}, and illustrated that the full SCS may not hold with large antenna array and wide bandwidth. That means the application of algorithms in \cite{Barbotin2011,Prasad2015} will be limited.
In this work, we focus on studying channel estimation algorithm for massive MIMO systems with large antenna array, where the full SCS doesn't often exist, but the channels in each cluster have SCS.

Pertaining to the aforementioned scenario, a message passing based channel estimation algorithm is proposed in this paper, which exploits the un-known SCS information by leveraging the cluster property of Dirichlet process (DP). By assuming that channels with SCS property share a precision vector, the unknown SCS information can be automatically learned by using the Dirichlet process.
Generally speaking, the proposed algorithm is based on the following two techniques.

(1) {\textit{Dirichlet process mixture.}} In the context of Bayesian non-parametric methods, DP mixture models \cite{Antoniak1974,Sethuraman1994} have been studied for more than three decades, and have been used in multi-target tracking \cite{Fox2006}, image segmentation \cite{Orbanz2008}, direction of arrival (DoA) estimation \cite{LuWang2016}, and many other scenarios. As in in \cite{Qi2008}, we also impose a DP prior over the sparse Bayesian learning (SBL) \cite{Tipping2001}, denote as DP-SBL, and apply such model in the channel estimation of massive MIMO system. 

(2) {\textit{Factor graph and message passing.}} Due to its remarkable performance, factor graph and message passing inference technology \cite{Kschischang2001} has been widely used in the design of wireless communication systems \cite{Wu2016,Schniter2011,Wu2014,OurSPL}. Since each of the message passing rules, e.g., belief propagation (BP) \cite{Kschischang2001}, mean field (MF) \cite{Winn2005}, expectation propagation (EP) \cite{Minka2001}, have their specialities, a method that combines BP, MF or EP as a unified framework on a same factor graph has been proposed \cite{Riegler2013}, \cite{OurSPL}, which keeps the virtues but avoids their respective drawbacks. In this paper, the DP-SBL model is built on factor graph and combined BP-MF message passing, while some messages are handled by the recently developed generalized approximate message passing (GAMP) to reduce the complexity \cite{Rangan2011},\cite{ourSBL}. Compared to the variational Bayesian (VB) method in literature \cite{Blei2006} \cite{Qi2008}, the proposed algorithm can reduce the complexity significantly.

In summary, the proposed channel estimation algorithm have the following distinctive features:
$(1)$ It utilizes the sparsity of the channel impulse response (CIR), and the feature that antennas in massive MIMO system can be grouped into clusters according to their SCS property. By the adoption of DP-SBL model, the SCS information can be automatically learned, thus channels with SCS can be estimated jointly.
$(2)$ The DP mixture is modeled and derived based on the factor graph and combined message passing, which can significantly reduce the complexity. Simulation results show that, the proposed SCS-exploiting channel estimation algorithm shows significant performance gain and robustness over other methods in literature.


The remainder of this paper is organized as follow. The transmission model and channel model of the MIMO-OFDM system is described in Section II. In Section III, we present the DP mixture and the probabilistic model. The message computation, schedule and the complexity comparison of the proposed message passing based algorithm are detailed in Section IV. Numerical results are provided in Section V.

\textit{Notation}- Boldface lower-case and upper-case letters denote vectors and matrices, respectively. Superscripts  $(\cdot)^*$ and $(\cdot){\tra}$ represent conjugation and transposition, respectively. The expectation operator with respect to a density $g(x)$ is expressed by $\left\langle f(x) \right\rangle_{g(x)} = \int f(x) g(x) dx / \int g(x') dx' $. The probability density function (pdf) of a complex Gaussian distribution with mean {$\hat x$} and variance $\nu_x$ is represented by $\CN(x;{\hat x},\nu_x)$. The pdf of Gamma distribution with shape parameter $a$ and scale parameter $b$ is denoted as $\Ga(x;a,b)$, and beta distribution with shape parameter $a, b$ is denoted as $\Be(x;a,b)$. The gamma and digamma function are represented by $\Gamma(x)$ and $\Psi(x)$ respectively. The relation $f(x)=cg(x)$ for some positive constant $c$ is written as $f(x)\propto g(x)$. We use the $\text{Diag}(\bx)$ to transform the vector $\bx$ into a diagonal matrix with the entries of $\bx$ spread along the diagonal.
\section{System Model}
\subsection{MIMO-OFDM Transmission Model}
Consider the uplink of a multiuser massive MIMO-OFDM system consisting of $U$ users, each of which equipped with one antenna, and a receiver equipped with $M$ antennas. To combat the inter symbol interference, the users are modulated by OFDM with $N_{\text{T}}$ subcarriers. The transmitted symbols by the $u$th user in frequency domain are denoted by $\bx_u = [x_u(1);...; x_u(N_{\text{T}})]\tra$. Among the $N_{\text{T}}$ subcarriers, $N$ uniformly spaced subcarriers are selected for all the users to transmit pilot signals, with $\mP_u$ represents  the indices set of pilot-subcarriers of user $u$. As in \cite{OurSPL}, we assumes that $\cup \mP_u=\emptyset$, and when a pilot-subcarrier is employed by a user, the remaining users do not use it to transmit any signal. The received signal $\by^{(m,u)}\in \mathbb{C}^{N\times 1}$ by the $m$th receive antenna from the $u$user can be written as
\begin{eqnarray}
\by^{(m,u)}&=&\bX^{(u)}\bh^{(m,u)}+n^{(m)},\label{eq:recv1}
\end{eqnarray}
where $\bX^{(u)}=\text{Diag}(\bx_u)\in \mathbb{C}^{N\times N}$ stands for the diagonalized pilot symbols of $u$th user, $\bh^{(m,u)}$ represents the vector of frequency-domain channel weight between the $u$th user and the $m$th receive antenna, $\bn^{(m)}\in \mathbb{C}^{N\times 1}$ represents the additive white Gaussian noise (AWGN) with zero mean and variance $\lambda^{-1}\boldsymbol{I}$.
Since users are independent to each other and $\cup \mP_u=\emptyset$, here we consider only one user without loss of generality.
In the rests of this paper, we will drop the script $u$ for convenience, then the receive model in \eqref{eq:recv1} becomes the simplified form
\begin{eqnarray}
\by^{(m)}=\bX\bh^{(m)}+\bn^{(m)}.\label{eq:recv2}
\end{eqnarray}

\subsection{Spatial Channel Model} \label{sec:Chnl}
It is known that most wireless channels can be modeled as multi-path channels with large delay spread and very few significant paths as scatterers are sparsely distributed in space. This makes the CIR sparse \cite{Bajwa2009, Minn2000}. Thus, for each transmit-receive link, we need only estimate a few significant multi-path channel gains, which has the potential to reduce the pilot overhead substantially.
Following \cite{Masood2015}\cite{Prasad2014}, we also build the frequency channel weight $\bh^{(m,u)}$ on tapped delay line model
\begin{eqnarray}
\bh^{(m,u)}=\bF^{(u)}\balpha^{(m,u)}, \label{eq:CFR1}
\end{eqnarray}
where $\bF^{(u)}\in \mathbb{C}^{N\times L}$ represents the truncated Fourier matrix formed by selecting the $\mP_u$ rows and the first $L$ columns from the discrete Fourier transform (DFT) matrix, $\balpha^{(m,u)}\in \mathbb{C}^{L\times 1}$ denotes the $L$-taps sparse channel between the $u$th user and $m$th receive antenna.
As \eqref{eq:recv2}, equation \eqref{eq:CFR1} can also be simplified as
\begin{eqnarray}
\bh^{(m)}=\bF\balpha^{(m)}. \label{eq:CFR2}
\end{eqnarray}

Due to the physical properties of outdoor electromagnetic propagation, the CIR measured at different antennas of MIMO systems share a common support, i.e. the times of arrival (ToA) at different antennas are similar while the paths amplitudes and phases are distinct \cite{Barbotin2011}. An example of the SCS channel model for a $4\times 4$ section of an antenna array is shown in \cite[Fig.3]{Masood2015} and \cite[Fig.1]{Barbotin2011}. Since the degrees of freedom to estimated can be reduces with such SCS assumption, which can improve the channel estimation overhead. 

\begin{figure}[!t]
\centering
\includegraphics[width=0.4\textwidth]{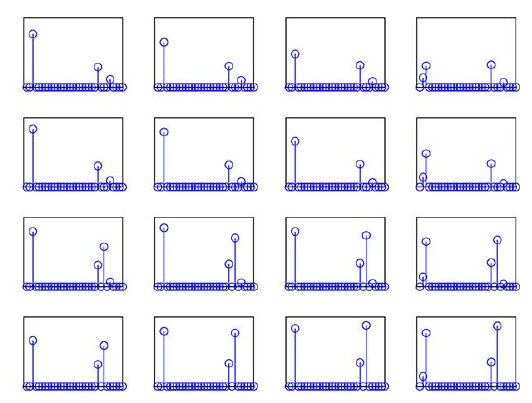}
\centering
\caption{Schematic diagram of an antenna array without full SCS property. Plots show the tap strengths on y-axis with respect to the tap locations on x-axis.}
\label{fig:ComSupt-UnCom}
\end{figure}

It is important to note that, the SCS assumptions only hold with respect to the channel bandwidth $Bw$ and the signal noise ratio (SNR) of the channel. One can assume that antennas with distance less than $d_{\text{max}}=\frac{C}{10\text{Bw}}$ share a common support \cite{Barbotin2011,Masood2015}. Authors in \cite{Masood2015} illustrate the relationship between the maximum resolvable distance ($d_{max}$) and the dimensions of the arrays for three different communication standards, with the distance between two adjacent antennas is assumed to be $d=\lambda/2$ where $\lambda$ is the signal wavelength. It can be seen that such SCS support may not hold with large bandwidth and large arrays. A schematic diagram of an antenna array without full SCS property is shown in Fig. \ref{fig:ComSupt-UnCom} \cite{Masood2015}.

To the authors' knowledge, there are lack of conclusive methods about the support pattern in massive MIMO. In this paper we set the channel of massive MIMO using a simple assumption: \textit{an antenna may have common support with its neighbors in probability $p$}. With such assumption, the channels can be grouped into several clusters, and antennas in each cluster have the property of common support. This model is more general than \cite{Barbotin2011}, i.e., when set $p=1$ the proposed channel model is equivalent to \cite{Barbotin2011}. The construction of aforementioned model is detailed in section \ref{sec:simu}.

\section{Unknown Sparse Common Support Using Sparse Bayesian Learning with Dirichlet Process}
To acquire the SCS information in the MIMO-OFDM systems, we resort to the sparse Bayesian learning (SBL) with Dirichlet process (DP) prior, e.g., DP-SBL, as in \cite{Qi2008} and \cite{LuWang2016}. In this section we first introduce the DP and SBL model briefly, and then present the DP-SPL model using probabilistic model and factor graph.
\subsection{SBL with Dirichlet Process Prior}
\subsubsection{Sparse Bayesian Learning Model}
Since equation \eqref{eq:recv2} is a typical SBL problem, here we employ a two-layer (2-L) hierarchical structure \cite{Tipping2001} that assumes a conditional prior pdf as
\begin{eqnarray}
p(\balpha^{(m)}|\bgamma)&=&\prod_l \CN(\alpha^{(m)}_l;0,\gamma_l^{-1}),~~~\forall m,\nonumber\\
p(\gamma_l)&=&\Ga(\gamma_l;c,d).\nonumber
\end{eqnarray}
The above equations imply that, all the sparse channel taps $\balpha^{(m)}, \forall m$, have the common hyper-prior, which is equivalent to the assumption of \cite{Prasad2015}. However, in this paper we are solving the problem that, the total $M$ channel may be grouped into several sets of clusters, and the common hyper-prior may only be appropriate within each cluster. Through the use of DP employed as the prior over $\bgamma$, we can simultaneously outperform the clustering and SBL.

\subsubsection{Dirichlet Process}
The Dirichlet process, denoted as $DP(\eta,G_0)$, is a measure on measure, and is parameterized by a positive scaling parameter $\eta$ and the base distribution $G_0$.
Assume each $\bgamma^{(k)}, k=1:K$, is drawn identically from $G$ and $G$ itself is a random measure drawn form a Dirichlet process.
\begin{eqnarray}
\bgamma^{(k)}&\sim& G,~~~ k=1:K,\nonumber\\
G&\sim& DP(\eta,G_0), \nonumber
\end{eqnarray}
where $\bgamma^{(k)}\sim G$ denotes that $\bgamma^{(k)}$ follows a $G$ distribution.
Since the explicit formulation of $G$ is unattainable, a definition of $G$ in terms of a stick-breaking construction was provided in \cite{Sethuraman1994}, as
\begin{eqnarray}
G=\sum\nolimits_{k=1}^\infty \omega_k\delta(\widetilde\bgamma^{(k)}) \label{eq:Dir_G}
\end{eqnarray}
with
\begin{eqnarray}
&&\omega_k=\pi_k\prod\nolimits_{i=1}^{k-1}(1-\pi_i),\label{eq:omega}\\
&&p(\widetilde\bgamma^{(k)})= G_0,\label{eq:G0}
\end{eqnarray}
where $\delta(\cdot)$ is the Dirac delta function, and parameter $\pi_k$ has the prior distribution $p(\pi_k|\eta)= \Be(\pi_k;1,\eta)$.
Known form \eqref{eq:pgamma} that, the base distribution $G_0$ is selected as Gamma distribution.
The infinite number of components in \eqref{eq:Dir_G} will inevitably results in an intractable complexity. In practice, the number of components is truncated to a relatively large number $K$. In this paper, $K$ is set to be the number of antennas $M$ without loss of generality \cite{LuWang2016}.

\subsection{Probabilistic Model and Factor Graph Representation}
Following the stick-breaking construction of DP mixture in \cite{Blei2006}, we introduce the assignment variables $z^{(m)}_k$, which can be defined by indicator function
$$ {\mathbbm{1}[z^{(m)}_k=1]} = \left\{
\begin{aligned}
1 & ~~~ &  z^{(m)}_k=1 \\
0 & ~~ &  z^{(m)}_k\neq 1
\end{aligned}
\right.
$$
which indicate the mixture components, i.e. $\{\widetilde\bgamma^{(k)}\}_{k=1:K}$, with which $\balpha^{(m)}$ is associated. Then the assignment vector $\bz^{(m)}$ has a multi-nomial distribution with a parameter set $\{\omega_k\}_{k=1:K}$, i.e.,
\begin{eqnarray}
p\big(\bz^{(m)}|\{\omega_k\}_{k=1:K}\big)=\text{Mult}(\{\omega_k\}_{k=1:K}).\nonumber
\end{eqnarray}
Using the deterministic relationship of $\omega_k$ and $\pi_k$ as in \eqref{eq:omega}, we can define the conditional distribution as
\begin{gather}
p(\bz^{(m)}|\bpi)~~~~~~~~~~~~~~~~~~~~~~~~~~~~~~~~~~~~~~~~~~~~~~~~~~~~~~~\nonumber\\
=\prod_{k=1}^K \Big(\pi_k \prod_{i=1}^{k-1}(1-\pi_i)\Big)^{\mathbbm{1}[z^{(m)}_k]}
\triangleq \prod_{k=1}^K f_{z_k}^{(m)}\big(z_k^{(m)},\bpi\big)~~\label{eq:fz1}\\
=\prod_{k=1}^{K} \pi_k^{\mathbbm{1}[z^{(m)}_k]} \prod_{i=k+1}^{K}(1-\pi_k)^{\mathbbm{1}[z^{(m)}_i]}
\triangleq f_{\bz}^{(m)}\big(\bz^{(m)},\bpi\big),\label{eq:fz2}
\end{gather}
with vectors $\bpi=[\pi_1,...,\pi_K]\tra$ and $\bz=[z_1^{(m)},...,z_K^{(m)}]\tra$.

The distribution of $\balpha^{(m)}$ conditional on $\bz^{(m)}$ and $\{\widetilde\bgamma^{(k)}\}_{k=1:K}$ can be expressed as
\begin{gather}
p\big(\balpha^{(m)}|\bz^{(m)},\widetilde\bgamma^{(k)},\forall k\big)=\prod_l\prod_{k} \CN\big(\alpha^{(m)}_l;0,1/\widetilde\gamma^{(k)}_l\big)^{\mathbbm{1}[z^{(m)}_k]}\nonumber \\
\triangleq \prod_l\prod_k f_{\tD_{k,l}}^{(m)}\big(\alpha^{(m)}_l ,z^{(m)}_k,\widetilde\gamma^{(k)}_l\big)\nonumber \\
\triangleq f_{\tbD}^{(m)}\big(\balpha^{(m)},\bz^{(m)},\{\widetilde\bgamma^{(k)}\}_{\forall k}\big).
\end{gather}

Following \cite{Qi2008}, we can also define conditional and prior distributions,
\begin{eqnarray}
p(\bpi|\eta)&=&\prod\nolimits_k p(\pi_k|\eta)=\prod\nolimits_k \Be(\pi_k;1,\eta)\nonumber\\
&\triangleq& \prod\nolimits_k f_{\pi_k}(\pi_k,\eta)\triangleq f_{\bpi}(\bpi,\eta),\nonumber\\
p(\eta)&=&\Ga(\eta;e,g)\triangleq f_{\eta}(\eta),\nonumber
\end{eqnarray}
and the mixture components $\widetilde\bgamma^{(k)}$ have the prior distribution
\begin{eqnarray}
p(\widetilde\bgamma^{(k)})&=&\prod\nolimits_l \Ga\big(\widetilde\gamma^{(k)}_l;c,d\big)
\triangleq \prod\nolimits_l f_{\gamma_l}^{(k)}\big(\widetilde\gamma^{(k)}_l\big)\nonumber\\
&\triangleq& f_{\bgamma}^{(k)}\big(\widetilde\bgamma^{(k)}\big).\label{eq:pgamma}
\end{eqnarray}
From the receive model presented in \eqref{eq:recv2}, the likelihood function of observation vector $\by^{(m)}$ can be written as
\begin{eqnarray}
p\big(\by^{(m)}|\bh^{(m)},\lambda\big)&=&\prod\nolimits_{n}\CN\big(y^{(m)}_n;x_n h^{(m)}_n,\lambda^{-1}\big)\nonumber \\
&\triangleq& f_{\tby}^{(m)}\big(\bh^{(m)},\lambda\big)
\triangleq \prod\nolimits_n f_{\ty_n}^{(m)}\big(h_n^{(m)}.\lambda\big)\nonumber
\end{eqnarray}
The deterministic constrains of $\bh^{(m)}$ and $\balpha^{(m)}$, as is shown in \eqref{eq:CFR2}, can be expressed as
\begin{eqnarray}
&&p(\bh^{(m)}|\balpha^{(m)})=\prod_n \delta\big(h^{(m)}_n-\bF^{(m)}_n\balpha^{(m)}\big)\nonumber \\
&&~~~\triangleq f_{\bdelta}^{(m)}\big(\bh^{(m)},\balpha^{(m)}\big)
\triangleq \prod\nolimits_n f_{\delta_n}^{(m)}\big(h_n^{(m)},\balpha^{(m)}\big).\nonumber
\end{eqnarray}
As in \cite{ourSBL}, we also define the prior of noise precision
\begin{eqnarray}
p(\lambda)=\Ga(\lambda;a,b)\triangleq f_{\lambda}(\lambda).\nonumber
\end{eqnarray}

From the receive model presented in \eqref{eq:recv2} and the SBL with DP prior model list above, the joint pdf of the collection of observed and unknown variables can be factorized as
\begin{eqnarray}
&&p(\by,\bh,\balpha,\bz,\widetilde\bgamma,\bpi,\eta,\lambda)\nonumber \\
&&=\prod\nolimits_m f_{\tby}^{(m)}\big(\bh^{(m)},\lambda\big)f_{\bdelta}^{(m)}\big(\bh^{(m)},\balpha^{(m)}\big)
f_{\bz}^{(m)}\big(\bz^{(m)},\bpi\big)\nonumber \\
&&~~~\times f_{\tbD}^{(m)}\big(\balpha^{(m)},\widetilde\bgamma^{(k)},\bz^{(m)},\forall k\big)
\prod\nolimits_k f_{\bgamma}^{(k)}\big(\widetilde\bgamma^{(k)}\big)\nonumber\\
&&~~~\times p\big(\bpi|\eta\big)p(\eta)f_\lambda(\lambda).
\end{eqnarray}

\begin{figure*}[!t]
\centering
\includegraphics[width=0.65\textwidth]{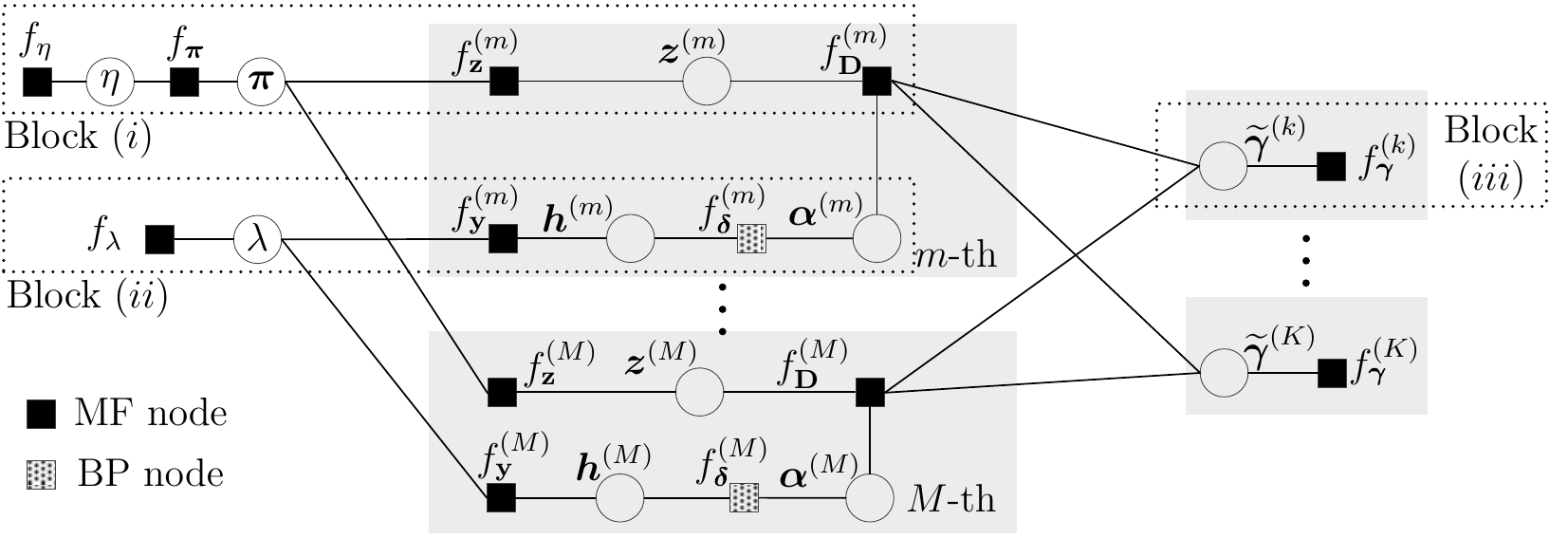}
\centering
\caption{Factor graph representations for the MIMO-OFDM system.}
\label{fig:Model}
\end{figure*}

The aforementioned factorization can be expressed in factor graph as depicted in Fig.\ref{fig:Model}.
For clarity, we group the factor graph into three functional blocks, labeled by Blocks $(i)-(iii)$ and marked in dashed boxes. Where Block $(i)$ represents the DP prior estimation, Block $(ii)$ denotes the estimation of hyper prior, and Block $(iii)$ represents the estimation of sparse channel taps and noise precision.

\section{Low Complexity Combined Message Passing Approach}
The message computation based on combined belief propagation (BP) and mean field (MF), message passing schedule and complexity comparison are presented in this section.
\subsection{Message Computation}
In this subsection, we detail the message computation in accordance with the three functional Blocks labeled in Fig.\ref{fig:Model}. Note that, if a forward message computation requires backward messages, we use the message in previous iteration by default.
\subsubsection{Messages Updating in DP Prior Estimation}
Assume the belief of $b(\alpha_l^{(m)})=\CN\big(\alpha_l^{(m)};\hat\alpha_l^{(m)},\nu_{\alpha_l}^{(m)}\big), \forall m, l$, are given from last iteration. We can compute the message $m_{f_{\tD_{k,l}}^{(m)}\to z_{k}^{(m)}}(z_{k}^{(m)})$ using MF rule,
\begin{align}
&m_{f_{\tD_{k,l}}^{(m)}\to z_{k}^{(m)}}(z_{k}^{(m)})=\exp\Big\{\left\langle \log f^{(m)}_{\tD_{k,l}}\right\rangle_{b(\alpha_l^{(m)})\prod\nolimits_k b(\widetilde\gamma_l^{(k)})}\Big\}\nonumber\\
&=\exp\Big\{{\mathbbm{1}[z^{(m)}_k]}
\left\langle\log \widetilde\gamma_l^{(k)}-\widetilde\gamma_l^{(k)}|\alpha_l^{(m)}|^2\right\rangle_{b(\alpha_l^{(m)}) b(\widetilde\gamma_l^{(k)})}\Big\}\nonumber\\
&=\exp\Big\{{\mathbbm{1}[z^{(m)}_k]}
\big(\big\langle\log \widetilde\gamma_l^{(k)}\big\rangle_{b(\widetilde\gamma_l^{(k)})}\nonumber\\
&~~~~~~~~~~~~~~~~~~~~~~~~-\big\langle\widetilde\gamma^{(k)}_l\big\rangle_{b(\widetilde\gamma^{(k)}_l)}(|\hat\alpha_l^{(m)}|^2+\nu_{\alpha_l}^{(m)})\big)\Big\},\nonumber
\end{align}
where $\big\langle\log \widetilde\gamma_l^{(k)}\big\rangle_{b(\widetilde\gamma_l^{(k)})}$ and $\big\langle\widetilde\gamma^{(k)}_l\big\rangle_{b(\widetilde\gamma^{(k)}_l)}$ denote the expectation of
$\log \widetilde\gamma_l^{(k)}$ and $\widetilde\gamma_l^{(k)}$ with respect to $b(\widetilde\gamma_l^{(k)})$, and their values are updated at \eqref{eq:blf_gama_m} and \eqref{eq:blf_loggama}.

With the factor node $f_{z_{k}}^{(m)}\big(z_k^{(m)},\bpi\big)$ defined in \eqref{eq:fz1} and the belief of $b(\bpi)$, later defined in \eqref{eq:blf_pi}, message $m_{f_{z_{k}}^{(m)}\to z_{k}^{(m)}}\big(z_{k}^{(m)}\big)$ can be updated by
\begin{align}
&m_{f_{z_{k}}^{(m)}\to z_{k}^{(m)}}\big(z_{k}^{(m)}\big)=\exp\Big\{\big\langle \log f_{z_{k}}^{(m)}\big\rangle_{b(\bpi)} \Big\}\nonumber\\
&=\exp\Big\{{\mathbbm{1}[z^{(m)}_k]} \big\langle \log\pi_k\sum\nolimits_{i=1}^{k-1}\log(1-\pi_i)\big\rangle_{b(\bpi)} \Big\}\nonumber\\
&=\exp\Big\{{\mathbbm{1}[z^{(m)}_k]}\big(\big\langle \log\pi_k \big\rangle_{b(\bpi)}+\sum\nolimits_{i=1}^{k-1}\big\langle \log(1-\pi_i)\big\rangle_{b(\bpi)}
\big) \Big\},\nonumber
\end{align}
where $\left\langle \log\pi_k \right\rangle_{b(\bpi)}
$ and $\left\langle \log(1-\pi_i)\right\rangle_{b(\bpi)}
$ represent the expectation of $\log\pi_k$ and $\log(1-\pi_i)$ with respect of the belief of $b(\bpi)$, and are updated in \eqref{eq:blf_logpi} and \eqref{eq:blf_log1_pi} respectively.
Then the belief of $b\big(z^{(m)}_k\big)$ can be obtained
\begin{eqnarray}
b\big(z^{(m)}_k\big)&=&m_{f_{z_{k}}^{(m)}\to z_{k}^{(m)}}\big(z_{k}^{(m)}\big)
\times \prod\nolimits_l m_{f_{\tD_{k,l}}^{(m)}\to z_{k}^{(m)}}\big(z_{k}^{(m)}\big)\nonumber\\
&=&\exp\big\{{\mathbbm{1}[z^{(m)}_k]}\times\hat E_{m,k}\big\}\nonumber
\end{eqnarray}
where $\hat E_{m,k}\triangleq \left\langle\log\pi_k \right\rangle_{b(\bpi)}+\sum\nolimits_{i=1}^{k-1}\left\langle \log(1-\pi_i)\right\rangle_{b(\bpi)}
+\sum\nolimits_l\big(\big\langle\log \widetilde\gamma_l^{(k)}\big\rangle_{b(\widetilde\gamma_l^{(k)})}
-\big\langle\widetilde\gamma^{(k)}_l\big\rangle_{b(\widetilde\gamma^{(k)}_l)}(|\hat\alpha_l^{(m)}|^2+\nu_{\alpha_l}^{(m)})\big)$.
After normalization, the expectation of $\big\langle{\mathbbm{1}[z^{(m)}_k]}\big\rangle_{b(z^{(m)}_k)}$ can be updated as
\begin{eqnarray}
\big\langle{\mathbbm{1}[z^{(m)}_k]}\big\rangle_{b(z^{(m)}_k)}=\frac{\exp \{\hat C_k\}}{\sum\nolimits_k \exp\{\hat C_k\}}\triangleq \hat\phi_{mk}.\label{eq:blf_phi}
\end{eqnarray}

With the definition of factor node $f_{\bz}^{(m)}\big(\bz^{(m)},\bpi\big)$ in \eqref{eq:fz2}, message $m_{f_{\bz}^{(m)}\to \bpi}(\bpi)$ can be updated by MF rule,
\begin{align}
&m_{f_{\bz}^{(m)}\to \bpi}(\bpi)=\exp\Big\{\left\langle \log f^{(m)}_{\bz}\right\rangle_{b(\bz^{(m)})}\Big\}\nonumber\\
&~~=\exp\Big\{\sum\nolimits_{k}\hat\phi_{mk}\log \pi_k + \sum\nolimits_{i=k+1}^{K-1}\hat\phi_{mi}\log(1-\pi_k)\Big\}.\nonumber
\end{align}

By the factor node, $f_{\bpi}(\bpi,\eta)=\prod\nolimits_k \Be(\pi_k;1,\eta)$, message $m_{f_{\bpi}\to \bpi}(\bpi)$ can be get by MF rule, i.e.,
\begin{eqnarray}
m_{f_{\bpi}\to \bpi}(\bpi)&=&\exp\left\{\left\langle \log f_{\bpi}\right\rangle_{b(\eta)}\right\}\nonumber\\
&=&\exp\big\{(\hat\eta-1)\sum\nolimits_k \log(1-\pi_k)\big\},\nonumber
\end{eqnarray}
where $\hat\eta$ denotes the expectation of $\eta$ with respect to $b(\eta)$, and is updated in \eqref{eq:blf_eta}.
Then the belief of $b(\bpi)$ can be get by
\begin{eqnarray}
b(\bpi)&=&m_{f_{\bpi}\to \bpi}(\bpi)\times \prod\nolimits_m m_{f_{\bz}^{(m)}\to \bpi}(\bpi)\nonumber\\
&=&\prod\nolimits_k\exp\Big\{\sum\nolimits_m \hat\phi_{mk}\log\pi_k \nonumber\\
&&~~+\Big[\sum\nolimits_m\sum\nolimits_{i=k+1}^{K}\hat\phi_{mi}+\hat\eta-1\Big] \log(1-\pi_k)\Big\}\nonumber\\
&=&\prod\nolimits_k \pi_k^{\tau_{k}^1-1}(1-\pi_k)^{\tau_{k}^2-1},\label{eq:blf_pi}
\end{eqnarray}
where
$\tau_{k}^1=\sum\nolimits_m \hat\phi_{mk}+1$ and $\tau_{k}^2=\sum\nolimits_m\sum\nolimits_{i=k+1}^{K} \phi_{mi}+\hat\eta$.
So the expectation of $\log \pi_k$ and $\log (1-\pi_k)$ with respect to the belief of $b(\bpi)$, can be get by, \cite{Qi2008},
\begin{eqnarray}
\left\langle\log \pi_k\right\rangle_{b(\bpi)}=\Psi(\tau_{k}^1)-\Psi(\tau_{k}^1+\tau_{k}^2)\label{eq:blf_logpi}\\
\left\langle\log (1-\pi_k)\right\rangle_{b(\bpi)}=\Psi(\tau_{k}^2)-\Psi(\tau_{k}^1+\tau_{k}^2),\label{eq:blf_log1_pi}
\end{eqnarray}
where $\Psi$ denotes the digamma function, with definition $\Psi(x)=\frac{d}{dx}\ln \Gamma(x)$.

Then the message $m_{f_{\bpi}\to \eta}(\eta)$ from factor node $f_{\bpi}$ to variable node $\eta$ is updated by MF rule, which reads
\begin{eqnarray}
m_{f_{\bpi} \to \eta}(\eta)&=&\exp\big\{\left\langle\log f_{\bpi}\right\rangle_{b(\bpi)}\big\}\nonumber\\
&=&\eta^{K}\exp\Big\{(\eta-1)\sum\nolimits_k \left\langle\log(1-\pi_k)\right\rangle_{b(\bpi)}\Big\}\nonumber\\
&\propto& \eta^{K}\exp\Big\{\eta\sum\nolimits_k \left\langle\log(1-\pi_k)\right\rangle_{b(\bpi)}\Big\}.\nonumber
\end{eqnarray}
With its prior $f_\eta(\eta)=\Ga(\eta;e,h)$, here we calculate the belief of $b(\eta)$ as
\begin{eqnarray}
b(\eta)&\propto& m_{f_{\bpi} \to \eta}(\eta)\times f_\eta(\eta)\nonumber\\
&\propto&\eta^{K+e-1}\exp\Big\{-\eta\big(h-\sum\nolimits_k\left\langle \log(1-\pi_k)\right\rangle_{b(\bpi)}\big)\Big\},\nonumber
\end{eqnarray}
and the expectation of $\eta$ can be updated
\begin{eqnarray}
\hat\eta=\left\langle\eta\right\rangle_{b(\eta)}=\frac{K+e-1}
{h-\sum\nolimits_k\left\langle \log(1-\pi_k)\right\rangle_{b(\bpi)}}.\label{eq:blf_eta}
\end{eqnarray}

\subsubsection{Messages Updating in Hyper Prior Estimation}
With the updated beliefs $b\big(\alpha^{(m)}_l\big)$ and $b\big(z_k^{(m)}\big)$  message $m_{f_{\tD_{k,l}}^{(m)}\to \widetilde\gamma^{(k)}_l}\big(\widetilde\gamma^{(k)}_l\big)$
from factor node $f_{\tD_{k,l}}^{(m)}\big(\alpha^{(m)}_l,\widetilde\gamma^{(k)}_l,z^{(m)}_k\big)$  to variable node $\widetilde\gamma^{(k)}_l$ can be get using MF
\begin{eqnarray}
m_{f_{\tD_{k,l}}^{(m)}\to \widetilde\gamma^{(k)}_l}\big(\widetilde\gamma^{(k)}_l\big)
=\exp\Big\{\big\langle \log f_{\tD_{k,l}}^{(m)}\big\rangle_{b\big(\alpha^{(m)}_l\big)\prod\nolimits_k b(z_k^{(m)})}\Big\}\nonumber\\
=\exp\Big\{ \hat\phi_{mk}\big(\log \widetilde\gamma^{(k)}_l-\widetilde\gamma^{(k)}_l(|\hat\alpha_l^{(m)}|^2+\nu_{\alpha_l}^{(m)})\big)\Big\}~~\nonumber\\
=\big(\widetilde\gamma_l^{(k)}\big)^{\hat\phi_{mk}}\exp\Big\{-\hat\phi_{mk}\big(|\hat\alpha^{(m)}_l|^2+\nu_{\alpha_l}^{(m)}\big)\widetilde\gamma^{(k)}_l\Big\}.\nonumber
\end{eqnarray}

Then by the prior of $f_{\gamma_l}^{(k)}\big(\widetilde\gamma_l^{(k)}\big)\propto \big(\widetilde\gamma^{(k)}_l\big)^{c-1}\exp\big\{-d\widetilde\gamma^{(k)}_l\big\}$, the belief of $\widetilde\gamma^{(k)}_l$ can be updated
\begin{eqnarray}
b\big(\widetilde\gamma^{(k)}_l\big)&\propto& f_{\gamma_l}^{(k)}(\gamma^{(k)}_l)
\times\prod\nolimits_m
m_{f_{\tD_{k,l}}^{(m)}\to \widetilde\gamma^{(k)}_l}(\widetilde\gamma^{(k)}_l)\nonumber\\
&=&\big(\widetilde\gamma^{(k)}_l\big)^{\hat c_{kl}-1}\exp\Big\{-\widetilde\gamma^{(k)}_l \hat d_{kl} \Big\}.\label{eq:blf_gama}
\end{eqnarray}
where $\hat c_{kl}=\sum\nolimits_m \hat\phi_{mk}+c$ and $\hat d_{kl}=\sum\nolimits_{m} \hat\phi_{mk}(|\hat\alpha^{(m)}_l|^2+\nu_{\alpha_l}^{(m)})+d$.
Thus the expectation of $\big\langle\widetilde\gamma^{(k)}_l\big\rangle_{b(\widetilde\gamma^{(k)}_l)}$ and $\big\langle\log\widetilde\gamma^{(k)}_l\big\rangle_{b(\widetilde\gamma^{(k)}_l)}$ can be updated as
\begin{eqnarray}
\big\langle\widetilde\gamma^{(k)}_l\big\rangle_{b(\widetilde\gamma^{(k)}_l)}&=&\frac{\hat c_{kl}}{\hat d_{ml}},\label{eq:blf_gama_m}\\
\big\langle \log \widetilde\gamma^{(k)}_l \big\rangle_{b(\widetilde\gamma^{(k)}_l)}&=&\Psi\big(\hat c_{kl}\big)-\log\big(\hat d_{ml}\big).
\label{eq:blf_loggama}
\end{eqnarray}

\subsubsection{Messages Updating in Sparse Channel and Noise Precision Estimation}
Assume that messages $m_{f_{\delta_n}^{(m)}\to h_n^{(m)}}(h_n^{(m)})=\CN\big(h_n^{(m)};\hat p_n^{(m)},\nu_{p_n}^{(m)}\big)$ from factor node $f_{\delta_n}^{(m)}\big(h_n^{(m)},\balpha^{(m)}\big)$ to variable node $h_n^{(m)}$ is obtained previously, which is defined in \eqref{eq:msg_p}.
Then the product of messages $m_{f_{\delta_n}^{(m)}\to \alpha_l^{(m)}}\big(\alpha_l^{(m)}\big), \forall n\in[1:N]$ can be get by
\begin{eqnarray}
q(\alpha^{(m)}_l)=\CN\big(\alpha_l^{(m)};\hat q_l^{(m)},\nu_{q_l}^{(m)}\big), \label{eq:msg_q}
\end{eqnarray}
where
\begin{eqnarray}
\nu_{q_l}^{(m)}&=&\Big(\sum\nolimits_{n}\frac{|F_{nl}^{(m)}|^2}{\nu_{\theta_n}^{(m)}+ \nu_{p_n}^{(m)}}\Big)^{-1}, \label{eq:msg_q_v}\\
\hat q_l^{(m)}&=&\nu_{q_l}^{(m)}\sum\nolimits_{n}\hat s_n^{(m)} (F_{nl}^{(m)})^*+\hat\alpha_l^{(m)},\label{eq:msg_q_m}
\end{eqnarray}
with $\hat s_n^{(m)}$ denotes a intermediate variable, which is defined as
\begin{eqnarray}
\hat s_n^{(m)}\triangleq \frac{\hat\theta^{(m)}_n-\hat p_n^{(m)}}{\nu^{(m)}_{\theta_n}+ \nu_{p_n}^{(m)}},\label{eq:msg_s}
\end{eqnarray}
and variables $\hat\theta^{(m)}_n$, $\nu^{(m)}_{\theta_n}$ represent the mean and variance of message $m_{f_{\ty_n}^{(m)}\to h_n^{(m)}}\big(h_n^{(m)}\big)$, which can be found in \eqref{eq:msg_theta}.
Note that, the derivation of equations \eqref{eq:msg_p}, \eqref{eq:msg_q} and \eqref{eq:msg_s} can be found in our prior work \cite[Eqs. (29)-(33)]{ourSBL}, and will not be detailed here.

With the beliefs of $b(z_k^{(m)})$ and $\big\{b(\widetilde\gamma_l^{(k)}),\forall k\big\}$, defined in \eqref{eq:blf_phi} and \eqref{eq:blf_gama}, message $m_{f_{\tbD_l}^{(m)}\to \alpha_l^{(m)}}(\alpha_l^{(m)})$ is computed by MF rule, which yields
\begin{eqnarray}
&&m_{f_{\tbD_l}^{(m)}\to \alpha_l^{(m)}}\big(\alpha_l^{(m)}\big)=\exp\Big\{\left\langle \log f_{\tbD_l}^{(m)}\right\rangle_{\prod\nolimits_k b(z_k^{(m)}) b(\widetilde\gamma_l^{(k)})}\Big\}\nonumber\\
&&~~=\exp\Big\{\sum\nolimits_k \hat\phi_{mk} \left\langle \log \widetilde\gamma^{(k)}_l-\widetilde\gamma^{(k)}_l|\alpha_l^{(m)}|^2 \right\rangle_{\prod\nolimits_k b(\widetilde\gamma^{(k)})}\Big\}\nonumber\\
&&~~\propto \CN\Big(\alpha_l^{(m)};0,\big(\sum\nolimits_k \hat\phi_{mk}\big\langle\widetilde\gamma^{(k)}_l\big\rangle_{b(\widetilde\gamma^{(k)}_l)}\big)^{-1}\Big).\nonumber
\end{eqnarray}
Then the belief of $\alpha_l^{(m)}$ is updated
\begin{eqnarray}
b(\alpha_l^{(m)})&\propto& m_{f_{\tbD_l}^{(m)}\to \alpha_l^{(m)}}\big(\alpha_l^{(m)}\big)\times q(\alpha^{(m)}_l)\nonumber\\
&\triangleq&\CN\big(\alpha_l^{(m)};\hat\alpha_l^{(m)},\nu_{\alpha_l}^{(m)}\big),\nonumber
\end{eqnarray}
with
\begin{eqnarray}
\nu_{\alpha_l}^{(m)}&=&\Big(\sum\nolimits_k \hat\phi_{mk}\big\langle\widetilde\gamma^{(k)}_l\big\rangle_{b(\widetilde\gamma^{(k)}_l)}+\big(\nu_{q_l}^{(m)}\big)^{-1}\Big)^{-1}, \label{eq:blf_alpha_v}\\
\hat\alpha_l^{(m)}&=&{\nu_{\alpha_l}^{(m)}\hat q_l^{(m)}}/{\nu_{q_l}^{(m)}}.\label{eq:blf_alpha_m}
\end{eqnarray}

With the GAMP method proposed in \cite{Rangan2011}, message $m_{f_{\delta_n}^{(m)}\to h_n^{(m)}}\big(h_n^{(m)}\big)$ can be updated by
\begin{eqnarray}
m_{f_{\delta_n}^{(m)}\to h_n^{(m)}}\big(h_n^{(m)}\big)=\CN\big(h_n^{(m)};\hat p_n^{(m)},\nu_{p_n}^{(m)}\big),\label{eq:msg_p}
\end{eqnarray}
where
\begin{eqnarray}
\nu_{p_n}^{(m)}&=&\sum\nolimits_{l}|F_{nl}^{(m)}|^2\nu_{\alpha_l}^{(m)}\label{eq:msg_p_v}\\
\hat p_n^{(m)}&=&\sum\nolimits_{l}F_{nl}^{(m)}\hat\alpha_l^{(m)}-\hat s_n^{(m)} \nu_{p_n}^{(m)}.\label{eq:msg_p_m}
\end{eqnarray}
Then message $m_{f_{\ty_n}^{(m)}\to h_n^{(m)}}\big(h_n^{(m)}\big)$, form the observation node $f_{\ty_n}^{(m)}(\h_n^{(m)},\lambda)$ to variable node $h_n^{(m)}$, can be updated as
\begin{eqnarray}
m_{f_{\ty_n}^{(m)}\to h_n^{(m)}}\big(h_n^{(m)}\big)&=&\exp\{\big\langle \log f_{\ty_n}^{(m)}\big\rangle_{b(\lambda)}\}\nonumber\\
&\triangleq& \CN\big(h_n^{(m)};\hat\theta^{(m)}_n,\nu^{(m)}_{\theta_n}\big)\label{eq:msg_theta},
\end{eqnarray}
where $\hat\theta^{(m)}_n=y_n^{(m)}/x_n$ and $\nu^{(m)}_{\theta_n}=1/\big(\hat\lambda |x_n|^2\big)$.
Thus we can calculate the belief of $b(h_n^{(m)})$ as,
\begin{eqnarray}
b(h_n^{(m)})&\propto& m_{f_{\ty_n}^{(m)}\to h_n^{(m)}}\big(h_n^{(m)}\big)\times m_{f_{\delta_n}^{(m)}\to h_n^{(m)}}\big(h_n^{(m)}\big)\nonumber\\
&\triangleq&\CN\big(h_n^{(m)};\hat h_n^{(m)},\nu_{h_n}^{(m)}\big),\nonumber
\end{eqnarray}
where
\begin{eqnarray}
\nu_{h_n}^{(m)}&=&\Big(\big(\nu_{p_n}^{(m)}\big)^{-1}+\big(\nu^{(m)}_{\theta_n}\big)^{-1}\Big)^{-1},\label{eq:blf_h_v} \\
\hat h_n^{(m)}&=&\nu_{h_n}^{(m)}\Big( \hat p_n^{(m)}/\nu_{p_n}^{(m)}+ \hat\theta^{(m)}_n/\nu^{(m)}_{\theta_n}\Big).\label{eq:blf_h_m}
\end{eqnarray}
Then the expectation of noise precision can be updated by,
\begin{eqnarray}
\hat\lambda=\frac{NM}{\sum\nolimits_{n,m} \big\langle |y_n^{(m)}-x_n h_{n}^{(m)}|^2\rangle_{b(h_{n}^{(m)})}}.\label{eq:Lmd}
\end{eqnarray}
Detailed derivation of \eqref{eq:Lmd} can be found in \cite{ourSBL,OurSPL}.

\subsection{Message Passing Schedule}
The factors in Fig. \ref{fig:Model} are vary densely connected and thus there are a multitude of different options for message passing scheduling. To improve the clarity, we summarize the schedule and the corresponding message updating of the proposed algorithm as \textbf{Algorithm 1}.

\begin{algorithm}
\caption{The Proposed Channel Estimation Algorithm}\label{alg:BPMFLC}
\begin{algorithmic}[1]
\State Initialize $\hat\phi_{mk}$, $\forall m,k$; $\nu_{p_n}^{(m)}$, $\hat p_n^{(m)}$, $\hat s_n^{(m)}$, $\forall m, n$;
$\big\langle\log \pi_k\big\rangle_{b(\bpi)}$, $\big\langle\log (1-\pi_k)\big\rangle_{b(\bpi)}$, $\forall k$; 
$\hat\eta$ and $\hat\lambda$.
\For{$t=1\to T $}
\State $\forall m,l$: update $\nu_{q_l}^{(m)}$ and $\hat q_l^{(m)}$ by \eqref{eq:msg_q_m} and \eqref{eq:msg_q_v}.
\State $\forall m,l$: update $\hat\alpha_l^{(m)}$ and $\nu_{\alpha_l}^{(m)}$ by \eqref{eq:blf_alpha_m} and \eqref{eq:blf_alpha_v}.
\State $\forall k,l$: update $\big\langle\widetilde\gamma^{(k)}_l\big\rangle_{b(\widetilde\gamma^{(k)}_l)}$ and $\big\langle \log\widetilde\gamma^{(k)}_l\big\rangle_{b(\widetilde\gamma^{(k)}_l)}$ by \eqref{eq:blf_gama}, \eqref{eq:blf_loggama}.
\State $\forall m,k$: update $\hat\phi_{mk}$ by \eqref{eq:blf_phi}.
\State $\forall k$: update $\big\langle\log \pi_k\big\rangle_{b(\bpi)}$ and $\big\langle\log (1-\pi_k)\big\rangle_{b(\bpi)}$ by

\eqref{eq:blf_logpi}, \eqref{eq:blf_log1_pi} respectively.
\State update $\hat\eta$ by \eqref{eq:blf_eta}.
\State $\forall k$: update $\big\langle\log \pi_k\big\rangle_{b(\bpi)}$ and $\big\langle\log (1-\pi_k)\big\rangle_{b(\bpi)}$ again

by \eqref{eq:blf_logpi} and \eqref{eq:blf_log1_pi} respectively.
\State $\forall m,k$: update $\hat\phi_{mk}$ again by \eqref{eq:blf_phi}.
\State $\forall m,l$: update $\hat\alpha_l^{(m)}$ and $\nu_{\alpha_l}^{(m)}$ again by \eqref{eq:blf_alpha_m} and \eqref{eq:blf_alpha_v}.
\State $\forall m,n$: update $\hat s_n^{(m)}$ by \eqref{eq:msg_s}.
\State $\forall m,n$: update $\nu_{h_n}^{(m)}$ and $\hat h_n^{(m)}$ by \eqref{eq:blf_h_m} and \eqref{eq:blf_h_v}.
\State update $\hat\lambda$ by \eqref{eq:Lmd}.
\State $\forall m,n$: update $\nu_{p_n}^{(m)}$ and $\hat p_n^{(m)}$ by \eqref{eq:msg_p_m} and \eqref{eq:msg_p_v}.
\EndFor\  $t$ \label{alg:end1}
\end{algorithmic}
\end{algorithm}
Firstly, we initialize variables which are used before updating. Since the clustering information are unknown for the receiver, we firstly set $\hat\phi_{mk}=1/K$. Other variables are initialized as $\nu_{p_n}^{(m)}=1$, $\hat p_n^{(m)}=0$, $\hat s_n^{(m)}=y_n^{(m)}\hat\lambda$, $\forall m,k$, $\big\langle\log \pi_k\big\rangle_{b(\bpi)}=\big\langle\log (1-\pi_k)\big\rangle_{b(\bpi)}=1/K$, $\hat\eta=1$, and $\hat\lambda=1$.

Then messages are updated iteratively and sequentially, until the maximum iteration number $T$ is reached, as shown in lines 2-16. Within each line, the messages are simultaneously computed, for all $n\in [1:N], m\in [1:M]$ and $k\in [1:K]$.

\begin{table}[!t]
\centering
\caption{Complexity comparison}
\centering
\includegraphics[width=0.49\textwidth]{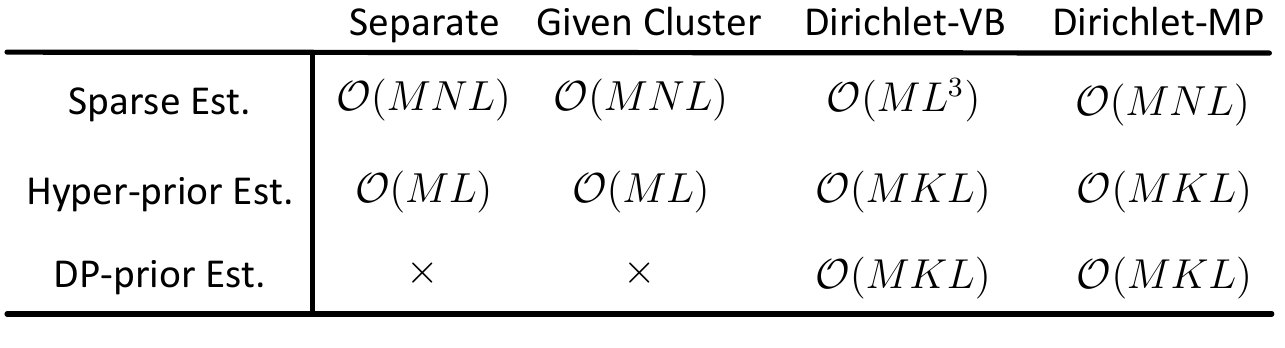}
\centering
\label{Tab:Complexity}
\end{table}
\subsection{Comparison of Computational Complexity}
As demonstrated in Section III, the proposed algorithms can be partitioned into three functional blocks:
DP-prior estimation (Block $i$), hyper-prior estimation (Block $ii$) and sparse channel and noise precision estimation (Block $iii$), which are respectively denoted as \textit{Sparse Est, hyper-prior Est,} and \textit{DP-prior Est}.
In the following, we compare the complexity of the mentioned algorithms in accordance with such partitions.
We use ``Dirichlet-MP'' to represents the proposed estimator based on DP-SBL and message passing, ``Dirichlet-VB'' to represents the adoption of variational Bayesian (VB) method in \cite{LuWang2016} and \cite{Qi2008} in the DP-SBL model, ``Separate'' to denotes the sparse Bayesian learning method proposed in \cite{ourSBL} which does not exploit the SCS property. The performance of the estimator with given SCS knowledge is also included for reference, which is denoted as ``GivenCluster''.

For the ``Separate'' and ``GivenCluster'', they have no DP-prior, and there are only $\mO(MNL)$ messages to be updated for the \textit{Sparse Est} and $\mO(ML)$ messages for the \textit{hyper-prior Est}. Since only several basic operations are need for each message updating, so the complexity of this two methods is $\mO(MNL)$.
Notice from the calculation of $b(\widetilde\bgamma^{(k)})$, $b(\bz^{(m)})$, the updating of \textit{DP-prior Est} and \textit{hyper-prior Est} for the Dirichlet-based algorithms (include ``Dirichlet-MP/VB'') require a complexity of $\mO(MKL)$. Due to the matrix inversion involved, ``Dirichlet-VB'' has a complexity of $\mO(ML^3)$ for the \textit{sparse Est} , while ``Dirichlet-MP'' has a complexity of $\mO(MNL)$ for this part. So the overall complexity for ``Dirichlet-VB'' is $\mO(MKL+ML^3)$ , and for ``Dirichlet-MP'' is $\mO(MKL+MNL)$.
The aforementioned complexity comparison is summarized in Table \ref{Tab:Complexity}.

\begin{figure}[!t]
\centering
\includegraphics[width=0.49\textwidth]{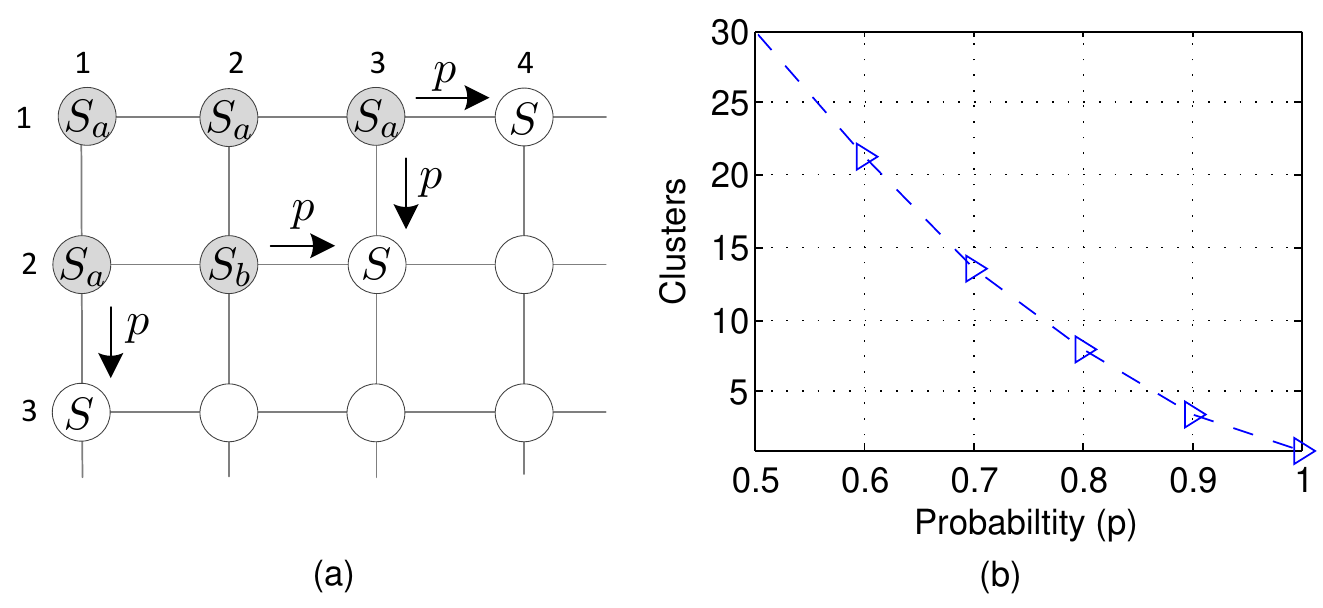}
\centering
\caption{(a) The creating of SCS model in massive MIMO system.
(b) The number of clusters grouped by (a) with different $p$, in a $10\times 10$ array. }
\label{fig:CreatCluster}
\end{figure}

\begin{figure}[!t]
\centering
\includegraphics[width=0.49\textwidth]{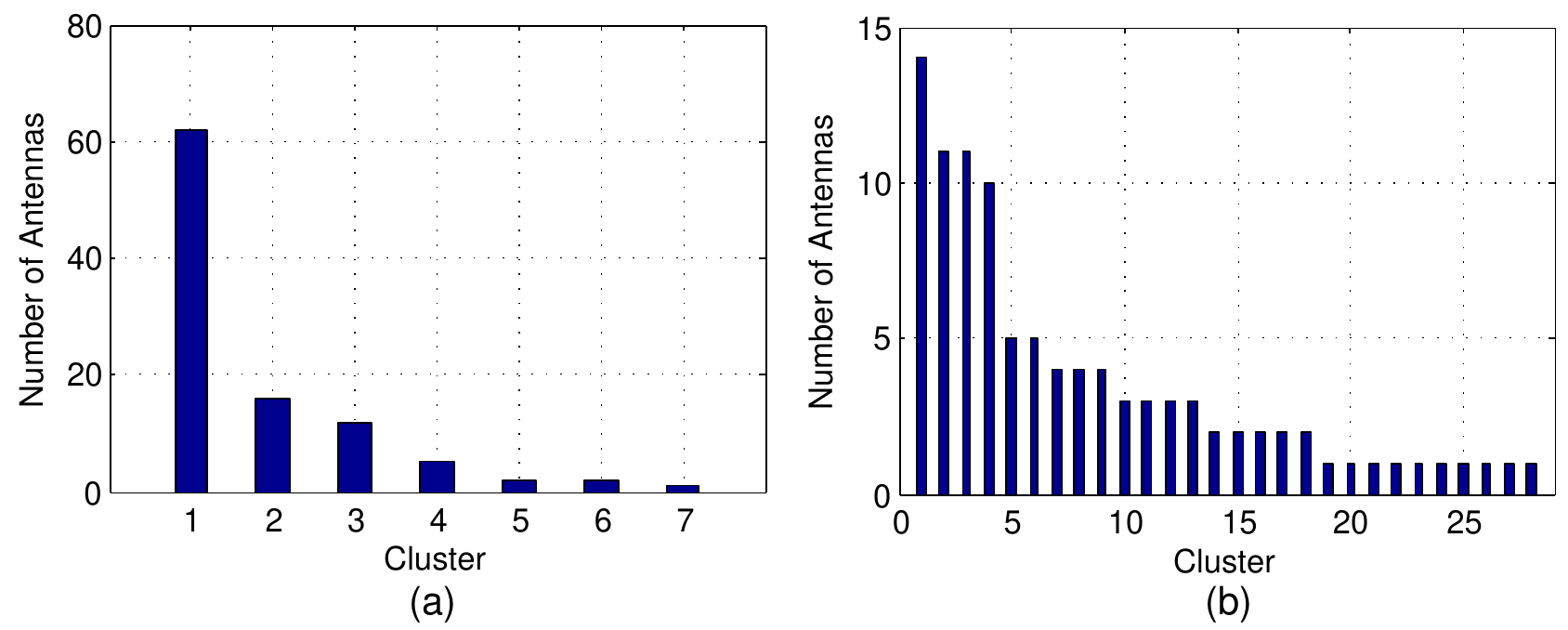}
\centering
\caption{Two typical SCS channel instances created by Fig. \ref{fig:CreatCluster}.(a), with probability (a) $p=0.8$, and (b) $p=0.5$.}
\label{fig:ClusterNum}
\end{figure}

\section{Numerical Experiments}\label{sec:simu}
In this section, we verify the performance advantages of our proposed channel estimation scheme via simulation. Specially, we consider a MIMO-OFDM system with the simulation parameters given in Table.\ref{tab:para} and channel model represented in Subsection \ref{sec:ChnlMod}.

\subsection{Channel Model of the Massive MIMO System}\label{sec:ChnlMod}
Known form the analysis in Subsection \ref{sec:Chnl} that, antenna array of the massive MIMO system can be grouped into several clusters, and the antennas share a common support within each cluster. Since lack of theory or measurement, the number of clusters and the range of each cluster can not be modeled explicitly. Furthermore, such prior information about SCS has not used in the proposed algorithm. For simplicity,  simulations in this paper is based on the following simple SCS model: an antenna may have common support with its neighbors in probability $p$.

As shown in Fig.~\ref{fig:CreatCluster} (a), \textit{we firstly set the antenna $a_{1,1}$\footnote{Subscript $\{1,1\}$ denote the 1th row and 1th column of the antenna array.} has support pattern $S_a$, then the following antennas are modeled sharing the same sparse pattern with its left and upper neighbor with probability $p$ successively and independently.} Note that, for antennas in the 1th row/column, only their left/upper neighbor are considered. After aforementioned process, the antenna array can be grouped into several clusters. The number of clusters and the size of each cluster are determined by probability $p$. As shown in Fig. \ref{fig:CreatCluster} (b), an array with $10\times 10$ antennas can be grouped into about 8 clusters with $p=0.8$, and are grouped into only one cluster when $p=1$, which is equivalent to the full SCS assumption in \cite{Barbotin2011,Prasad2015}. Fig. \ref{fig:ClusterNum} demonstrate two typical instances of the SCS channel created by Fig.~\ref{fig:CreatCluster} (a), with probability (a) $p=0.8$, and (b) $p=0.5$. Fig. \ref{fig:ClusterNum} (a) shows that, when set $p=0.8$, the antenna array is grouped into 7 clusters, with the largest cluster has 61 elements, and the smallest cluster only has 1 element. As shown in Fig. \ref{fig:ClusterNum} (b), with $p=0.5$, the array is grouped into 28 clusters, and most of which has less than 5 elements.

\subsection{Performance Comparison of Various Estimators}
In this subsection, numerical simulations are conducted to evaluate the performance of the proposed algorithm in comparison with other reported ones.

\begin{table}[htb]
  \centering
  \begin{minipage}[t]{0.9\linewidth}
  \caption[table]{Parameters setting for the MIMO-OFDM system}
  \label{tab:para}
    \begin{tabular*}{\linewidth}{lp{16cm}}
      \toprule[1.5pt]
      {Antenna Array in BS} & $10\times 10$~~\\
      {Number of Subcarriers ($N_\text{Total}$)} & $512$~~\\
      {Channel Length ($L$)} & 64~~\\
      {Non-zeros Taps}  & {8}~~\\
      {Evenly Spaced Pilot Subcarriers ($N$)} & 24$\sim$40~~\\
      {Probability ($p$)}  & {0.5$\sim$1}~~\\
      \bottomrule[1.5pt]
    \end{tabular*}
  \end{minipage}
\end{table}

\begin{figure*}[!t]
\centering
\includegraphics[width=1\textwidth]{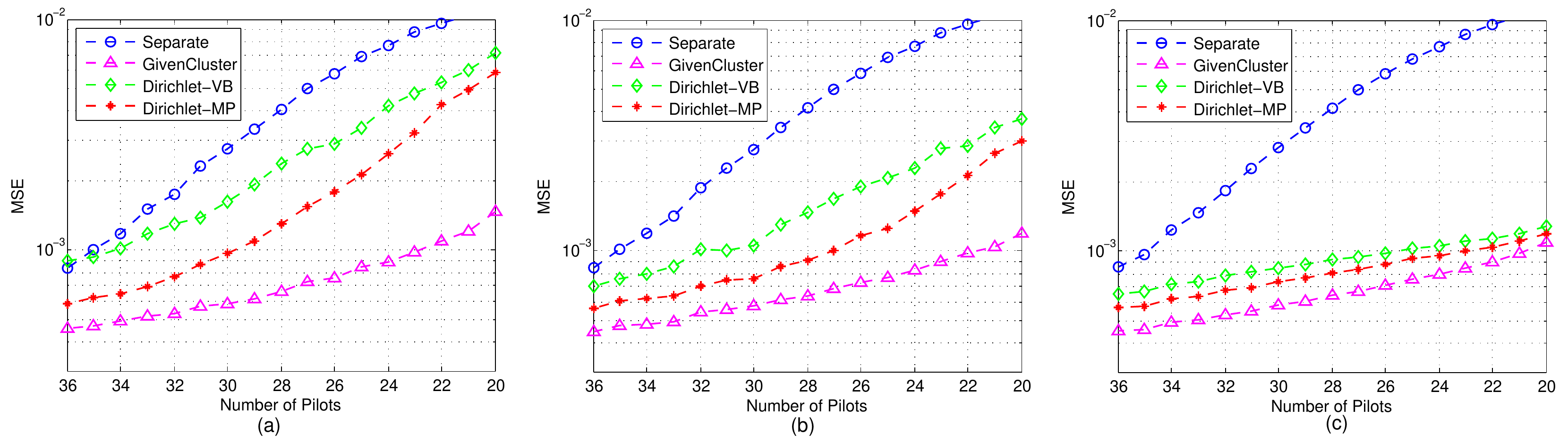}
\centering
\caption{MSE versus number of pilots, with SNR=10dB, and probability, (a) $p=0.8$, (b) $p=0.9$, (c) $p=1$.}
\label{fig:MSEvsPilots}
\end{figure*}

Fig. \ref{fig:MSEvsPilots} depicts mean-square-error (MSE) performance versus the number of pilots, with an signal-to-noise ratios (SNR) of 10dB. Note that, sub-figures Fig. \ref{fig:MSEvsPilots} (a)$\sim$(c) demonstrate the performance with different probability $p$, which determine the number of clusters with SCS.
Fig. \ref{fig:MSEvsPilots} (a)$\sim$(c) suggest that, 1) ``Separate'' keep invariant with different $p$, since the cluster information has not been considered, 2) ``GivenCluster'' has slight performance improvement with the increasing of cluster size, 3) ``Dirichlet-VB'' and ``Dirichlet-MP'' have similar performance with  ``GivenCluster'' for $p=1$, but exhibit some performance loss with $p=0.9$ and $p=0.8$. But compared to the existing ``Separate'' method, the Dirichlet-based methods (include ``Dirichlet-VB'' and ``Dirichlet-MP'') exhibit significant performance improvement. In other words, the Dirichlet-based methods could greatly reduce the channel estimation overhead.

\begin{figure*}[!t]
\centering
\includegraphics[width=1\textwidth]{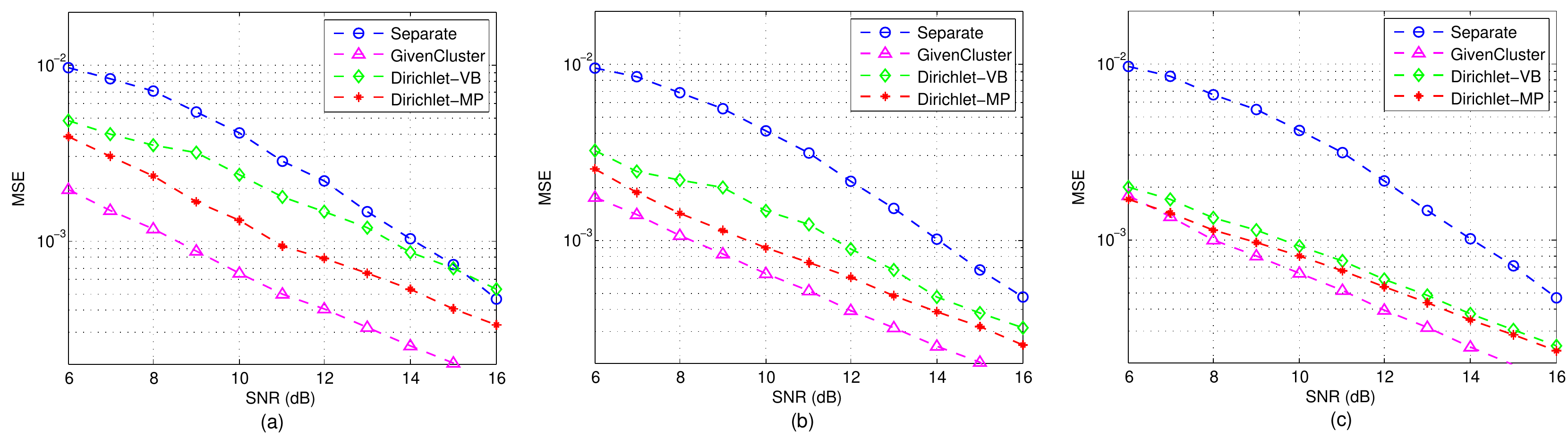}
\centering
\caption{Performance comparison in different SNR, with 28 pilots, and probability, (a) $p=0.8$, (b) $p=0.9$, (c) $p=1$.}
\label{fig:MSEvsSNR}
\end{figure*}

\begin{figure}[!t]
\centering
\includegraphics[width=0.4\textwidth]{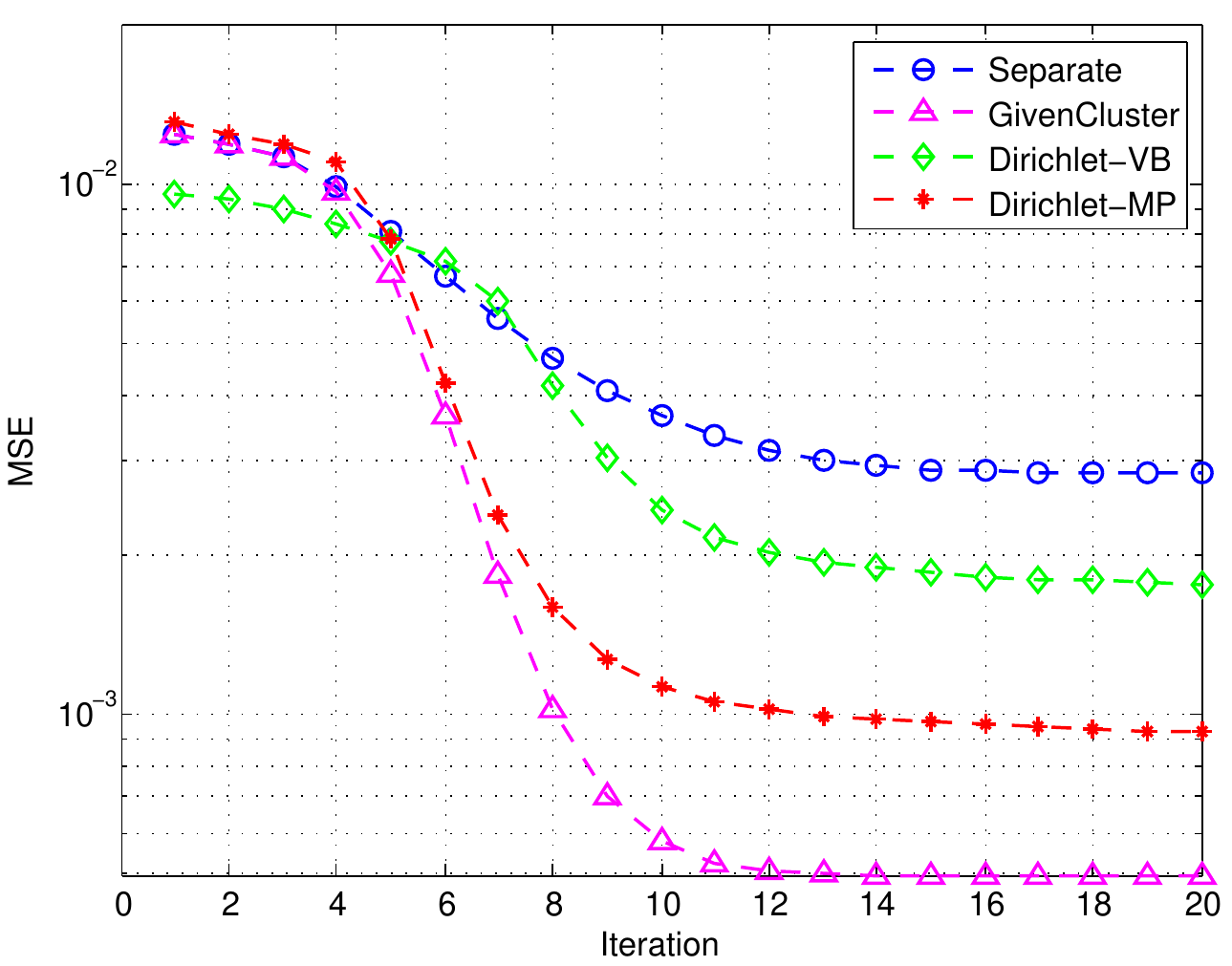}
\centering
\caption{MSE performance versus iteration, with SNR=10dB and 28 pilots.}
\label{fig:MSEvsIter}
\end{figure}

In Fig. \ref{fig:MSEvsSNR} (a)$\sim$(c), the MSE performance of the various algorithms is shown over the SNRs, where all algorithms run 20 iterations and the number of pilots employed fixed at 28. As Fig. \ref{fig:MSEvsPilots}, we also list the curves with different $p$. We observe that the proposed ``Dirichlet-MP'' approaches the ``GivenCluster'' while outperforms other methods.
Fig.\ref{fig:MSEvsIter} illustrates the MSE performance of the algorithms, operating at SNR = 10dB and $N = 26$, versus iteration index. It can be seen that our proposed algorithms have almost the same convergence speed with the ``GivenCluster''.

\begin{figure}[!t]
\centering
\includegraphics[width=0.4\textwidth]{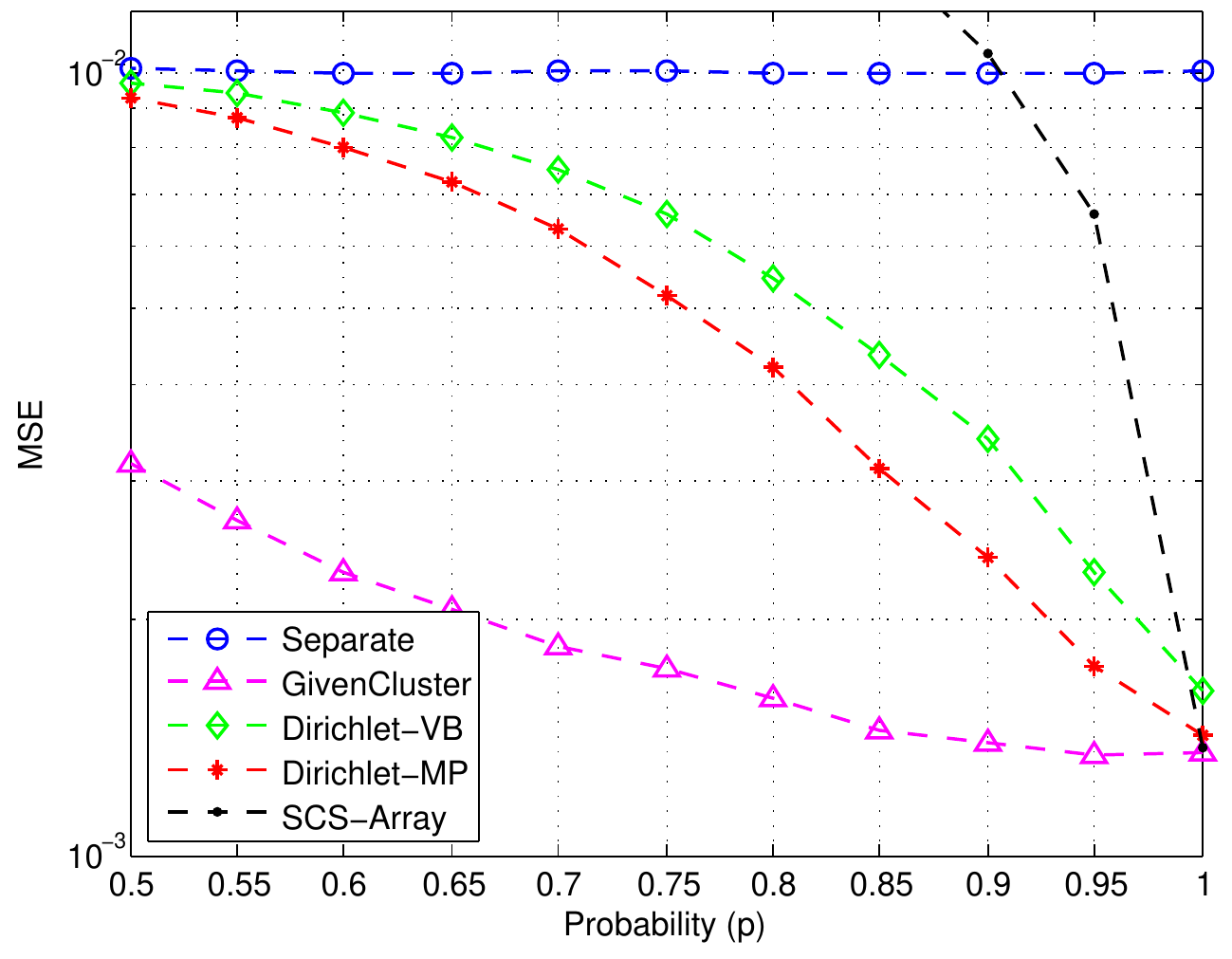}
\centering
\caption{MSE performance versus probability $p$, with SNR=8dB and 24 pilots.}
\label{fig:MSEvsProb}
\end{figure}

In Fig.~\ref{fig:MSEvsProb} shows MSE performance versus different probability $p$, with SNR=8dB and $N=24$.
It shows that the performance of ``Separate'' is fixed with different $p$, since no SCS property is utilized, ``GivenCluster'' exhibits better performance with the increasement of $p$, since larger $p$ indicates less clusters and larger cluster size.
{The ``Dirichlet-VB'' and the proposed ``Dirichlet-MP'' also show better performance with larger $p$, but their performance deteriorate with the decrease of $p$, even may slightly worse than ``Separate'' when $p\leq0.5$. We can explain such interesting result as follows.
Small $p$ indicates more clusters and fewer antennas within each cluster. As shown in Fig \ref{fig:ClusterNum} (b), the antenna array is grouped into 28 clusters with $p=0.5$, and most of the clusters only has 1-3 elements. Such dense clusters may lead to errors in the grouping of antennas for the Dirichlet-based algorithms, i.e., channels have no SCS property be grouped into one cluster, and such errors will lead to performance loss.}
{Note that, we add a new curve in Fig.~\ref{fig:MSEvsProb}, denotes the estimator which roughly assume the whole array have the SCS property, and is denoted as ``SCS-Array''. It shows that, the ``SCS-Array'' have the same performance with ``GivenCluster'' with $p=1$, but deteriorate rapidly when $p<1$. In the other hand, compared to ``SCS-Array'' the robustness can be significantly improved with the SCS-exploiting algorithm proposed in this paper.}

\section{Conclusion}
Massive MIMO systems provide substantial performance gains as compared to the traditional MIMO systems. However, these gains come with a huge requirement of estimating a large number of channels. In this paper we proposed a novel channel estimation algorithm, which utilize the fact that channels in a large antenna array may be grouped into clusters according to the sparsity pattern. By the adoption of Dirichlet prior over SBL model, the proposed algorithm can automatically learn the SCS information, thus channels with SCS can be estimated jointly. Furthermore, the combined message passing is used to derive the Dirichlet process mixture, which significantly reduced the complexity. Simulations demonstrate that, the proposed algorithm shows significant performance gain compared to methods in literature.
\bibliographystyle{IEEEtran}
\bibliography{IEEEabrv,bibliography}
\end{document}